\documentclass[conference]{IEEEtran}
\IEEEoverridecommandlockouts
\usepackage{cite}
\usepackage{comment}
\usepackage{amsmath,amssymb,amsfonts}
\usepackage{nicefrac}
\usepackage{float}
\usepackage{algorithmic}
\usepackage{graphicx}
\usepackage{textcomp}
\usepackage{xcolor}
\usepackage{nicefrac}
\usepackage{csquotes}
\usepackage{booktabs}
\usepackage{mathtools}
\usepackage{tabularx}
\usepackage{xfrac}
\usepackage{bm}
\newcolumntype{L}{>{\raggedright\arraybackslash}X}

\def\BibTeX{{\rm B\kern-.05em{\sc i\kern-.025em b}\kern-.08em
    T\kern-.1667em\lower.7ex\hbox{E}\kern-.125emX}}
\begin{document}

\newcommand{\rita}[1]{{\color{orange}{\textbf{REVIEW:}[[{#1}]]}}}

\title{A Tsallis-Entropy Lens on Genetic Variation \\
}

\author{\IEEEauthorblockN{Margarita Geleta}
\IEEEauthorblockA{\textit{Dept. of Computer Science} \\
\textit{University of California, Berkeley}\\
Berkeley, CA, USA \\
geleta@berkeley.edu}
\and
\IEEEauthorblockN{Daniel Mas Montserrat}
\IEEEauthorblockA{\textit{Dept. of Biomedical Data Science} \\
\textit{Stanford University School of Medicine}\\
Stanford, CA, USA \\
dmasmont@stanford.edu}
\and
\IEEEauthorblockN{Alexander G. Ioannidis}
\IEEEauthorblockA{\textit{Dept. of Biomedical Data Science} \\
\textit{Stanford University School of Medicine}\\
Stanford, CA, USA \\
ioannidis@stanford.edu}
}

\maketitle

\begin{abstract}
We introduce an information-theoretic generalization of the fixation statistic, the Tsallis-order $\bm{q}$ F-statistic, $\bm{F_q}$, which measures the fraction of Tsallis $\bm{q}$-entropy lost within subpopulations relative to the pooled population. The family nests the classical variance-based fixation index $\bm{F_{\textbf{ST}}}$ at $\bm{q{=}2}$ and a Shannon-entropy analogue at $\bm{q{=}1}$, whose absolute form equals the mutual information between alleles and population labels. By varying $\bm{q}$, $\bm{F_q}$ acts as a spectral differentiator that up-weights rare variants at low $\bm{q}$, while $\bm{q{>}1}$ increasingly emphasizes common variants, providing a more fine-grained view of differentiation than $\bm{F_{\textbf{ST}}}$ when allele-frequency spectra are skewed. On real data (865 Oceanian genomes with 1,823,000 sites) and controlled genealogical simulations (seeded from 1,432 founders from Human Genome Diversity Project and 1000 Genomes panels, with 322,216 sites), we show that $\bm{F_q}$ in One-vs-Rest (OVR) and Leave-One-Out (LOO) modes provides clear attribution of which subpopulations drive regional structure, and sensitively timestamps isolation-migration events and founder effects. $\bm{F_q}$ serves as finer-resolution complement for simulation audits and population-structure summaries. % Our empirical $\bm{F_q}$ patterns in Polynesia, Micronesia, Melanesia, and Southeast Asia accord with genomic reconstructions of settlement, drift, and admixture in Oceania.

\end{abstract}


\section{Introduction}

When characterizing populations, demographies, or bottlenecks, or when simulating genotypes (i.e., single-nucleotide polymorphism sequences or SNPs) \cite{geleta2023deep, perera2022generative, battey2021visualizing, yelmen2021creating,montserrat2019class}, it is important to understand genetic differentiation \cite{jakobsson2013relationship, morrison2022fstruct, rosenberg2025mathematical}. SNP sequences are predominantly biallelic with each genetic marker able to be coded in most cases as having two possible states, often denoted as \emph{reference} and \emph{alternate} alleles or \textit{ancestral} and \textit{derived}. Several metrics derived from population genetics are commonly used to quantify differentiation or population structure; yet the field still relies primarily on statistics tied to variance in allele frequencies \cite{meirmans2011assessing}—most prominently Wright’s F-statistics family \cite{wright1949genetical, weir1984estimating, meirmans2011assessing}, in particular the fixation index $F_{\text{ST}}$, and related variants such as Nei's $G_{\text{ST}}$ \cite{nei1973analysis} and Hedrick's $G'_{\text{ST}}$ \cite{hedrick2005standardized}. An alternative is Jost's $D$ \cite{jost2008gst}, which derives from allelic diversity and measures differentiation independent of within-population diversity. These metrics emphasize common variants and can lose sensitivity when allele-frequency spectra are skewed \cite{jost2018differentiation, chao2015expected}---situations increasingly common in whole-genome sequencing datasets.

We unify variance- and information-centric views by defining a \emph{Tsallis-order $q$ F-statistic}, $F_q$, which measures the fraction of Tsallis $q$-entropy lost within subpopulations relative to the pooled population. The family recovers the classical heterozygosity-based $F_{\text{ST}}$ at $q{=}2$ and a Shannon analogue at $q{=}1$, whose absolute form equals mutual information between allele and population labels. 
Our empirical analyses of $F_q$ patterns in Polynesia, Micronesia, Melanesia, and Southeast Asia accord with genomic reconstructions of settlement, drift, and admixture in Oceania.

\section{Background and related work}

\subsection{Variance-Based Divergence}
The genetic diversity of a population is often measured by the expected heterozygosity, $H_e$, which is the probability that two randomly drawn alleles differ or, alternatively, it is twice the variance of a Bernoulli-coded allele for a biallelic locus with allele frequency $p$:
\begin{equation}\label{eq:heterozygosity}
H_e = 1 - p^2 - (1-p)^2 = 2p(1-p).
\end{equation}
Wright's fixation index $F_{\text{ST}}$ partitions diversity into within- and total components and remains foundational in population genetics \cite{wright1949genetical}. A common biallelic form under Hardy-Weinberg equilibrium within subpopulations compares the heterozygosity of the total population, $H_T$, to the mean heterozygosity across subpopulations, $H_S$ \cite{weir1984estimating, nei1973analysis}:
\begin{equation}\label{eq:fst}
F_{\text{ST}} = \frac{H_T - H_S}{H_T} = \frac{\mathrm{Var}_w(p)}{\bar{p}(1-\bar{p})}
\end{equation}
quantifying the fraction of genetic diversity lost due to population subdivision, where $\mathrm{Var}_w(p)=\sum_{k=1}^K w_k (p_k-\bar{p})^2$ is the weighted variance across $K$ subpopulations. 
Values near $0$ indicate identical allele frequencies across subpopulations, whereas values approaching $1$ require extreme conditions, such as complete fixation of alternate alleles in populations with close to null within-population diversity. 
Although intuitive, $F_{\text{ST}}$ emphasizes common alleles (a second-order statistic) and is a \emph{relative} measure reporting the proportion of  variance explained rather than the absolute magnitude of allele frequency differences. 

Jost introduced $D$ as an \emph{absolute} measure of allelic differentiation, derived from effective numbers of alleles, and designed to decouple differentiation from within-population heterozygosity \cite{jost2008gst}. 
Although $D$ also ranges from 0 to 1, intermediate values lack a direct coalescent or migration-based interpretation, unlike $\text{F}_{\text{ST}}$ which can, under strong assumptions, be an indicator of gene flow \cite{whitlock1999indirect}.

\subsection{Information-Theoretic Diversity Measures}
Shannon entropy $H$ quantifies the expected uncertainty of allele identity; its exponent $e^H$ is widely used as a diversity index in ecology \cite{chao2013entropy, sherwin2010entropy}. For a biallelic locus $X$ with allele probabilities $p$ and $1-p$; we define:
\begin{equation}\label{eq:shannon}
H_{X\sim\text{Bern}(p)}(X)=-p\log p - (1-p) \log (1-p)
\end{equation}
which quantifies how uncertain (or \enquote{surprising}) an allele draw is on average. Unlike heterozygosity, which is quadratic in allele frequency, Shannon entropy is a first-order statistic; it downweights extremely common alleles and remains sensitive when the minor-allele spectrum is skewed. In population genetics, \emph{Shannon differentiation} (normalized mutual information) formalizes a relative information-based divergence between subpopulations and the pool, with expectations under the finite-island model \cite{chao2015expected}. 

\emph{Tsallis $q$-entropy} generalizes Shannon while remaining concave for $\forall q>0$, enabling Jensen-type decompositions across the entire $q$-range \cite{tsallis2004nonextensive} and has been previously explored as a diversity measure on transcriptomic datasets \cite{derian2022tsallis}.

\section{Theory --  a Tsallis $q$-Entropy F-Statistic }

\subsection{Tsallis $q$-Entropy (Biallelic) and Notation}
Let $Y\in\{1,...,K\}$ be the population label with $P(Y{=}k)=w_k$, where $w_k>0$ and $\sum_k w_k=1$. Let $X_\ell\in\{0,1\}$ denote the allele at locus $\ell$. Conditional on $Y=k$, we write $X_\ell|Y{=}k\sim\text{Bern}(p_{k\ell})$, so marginally $X_\ell \sim \text{Bern}(\bar{p}_\ell)$. For a random variable $X$ with a Bernoulli probability mass function with parameter $p$, the Tsallis entropy of order $q$ is defined as:
\begin{equation}
S_{q}(X) = \frac{1 - \big(p^q + (1-p)^q\big)}{q-1}, \quad q>0,\ q\neq 1,
\end{equation}
with $S_1(X)=H(X)$ (Equation \ref{eq:shannon}) by using l'Hôpital rule at the limit  $q\rightarrow1$. Define per-locus totals:
\[
\begin{aligned}
S_q^{\text{total}}(\ell)&\coloneqq S_q(X_\ell)=S_q(\text{Bern}(\bar{p}_\ell)) \\
S_q^{\text{within}}(\ell)&\coloneqq S_q(X_\ell | Y)=\sum_{k=1}^K w_k\,S_q(\text{Bern}(p_{k\ell}))
\end{aligned}
\]

\subsection{Absolute and Relative Differentiation}
Define the \emph{absolute} $q$-differentiation (Jensen–Tsallis gap): 
\begin{equation}\label{eq:absolute}
\Delta_q(\ell) \coloneqq S_q^{\text{total}}(\ell) - S_q^{\text{within}}(\ell) \ge 0,
\end{equation}
and the \emph{relative} $q$-statistic per locus:
\begin{equation}\label{eq:relative}
F_q(\ell) \coloneqq \frac{\Delta_q(\ell)}{S_q^{\text{total}}(\ell)} \quad \text{(defined when } S_q^{\text{total}}(\ell)>0\text{)}.
\end{equation}

\subsection{Properties}
\paragraph{Non-Negativity and Bounds}
For $\forall q>0, \Delta_q(\ell)\geq 0$ and $0\leq F_q(\ell)\leq 1$. 

\emph{Proof}: $S_q$ is concave for $q{>}0$; by Jensen, $S_q(\text{Bern}(\bar{p}_{\ell})) \geq \sum_k w_k S_q(\text{Bern}(p_{k\ell}))$, hence $\Delta_q(\ell)\geq 0$. Since $\sum_k w_k S_q(\text{Bern}(p_{k\ell}))\geq 0$ and $S_q(\text{Bern}(\bar{p}_{\ell}))\geq \Delta_q(\ell)$, the ratio lies between $0$ and $1$.

\paragraph{Reductions to Canonical Measures ($q=2$ and $q=1$)}
For $q=2$, $S_2(\text{Bern}(p))=2p(1-p)$, which equals $H_e$ (Equation \ref{eq:heterozygosity}). A direct computation yields:
\begin{equation}
\Delta_2(\ell) = 2\bar{p}(1-\bar{p})-2\sum_k w_k p_{k\ell}(1-p_{k\ell})
= 2\,\mathrm{Var}_w(p_{k\ell}),
\end{equation}
Hence $F_2(\ell)=(H_T-H_S)/H_T=F_{\text{ST}}$ via Equation \ref{eq:fst}.

For $q=1$, recall the marginal distribution $P(X_\ell)=\bar{p}_\ell =\sum_{k=1}^K w_kp_{k\ell}$. Then:
\begin{equation}
\begin{aligned}
\Delta_1(\ell) &= 
H(X_\ell)  -\sum_{k=1}^K w_k H(X_\ell | Y=k)
\\&= 
H(X_\ell)-H(X_{\ell}|Y)=I(X_\ell;Y),
\end{aligned}
\end{equation}
the mutual information between the allele at locus $\ell$ and the population label (equivalently, the weighted Jensen–Shannon divergence). 
Thus, $F_1(\ell)=I(X_\ell;Y)/H(X_\ell)$ is \emph{Shannon differentiation}.

\subsection{One-vs-Rest $F_q$ (OVR): Per-Population Separation Within a Region}
For a focal population $c$ inside a macro-region $\mathcal{R}$, we define two groups at each locus $\ell$: group $A$ collects the set of haplotypes from $c$, and group $B$ represents the set of haplotypes pooled from $\mathcal{R}\setminus\{c\}$. We use equal group weights $w_A=w_B=\nicefrac{1}{2}$ so populations contribute as units rather than by sample size (we down-sample large populations per bootstrap replicate to stabilize variance). With per-group allele frequencies $p_{A\ell}, p_{B\ell}$ and pooled $\bar{p}_\ell=\nicefrac{1}{2}(p_{A\ell}+p_{B\ell})$, define:
\begin{equation}
\begin{aligned}
S_q^{\text{total}}(\ell)&=S_q(\text{Bern}(\bar{p}_\ell))\\
S_q^{\text{within}}(\ell)&=\nicefrac{1}{2}S_q(\text{Bern}(p_{A\ell}))+\nicefrac{1}{2}S_q(\text{Bern}(p_{B\ell}))
\end{aligned}
\end{equation}
and absolute and relative differentiation follow Equations \ref{eq:absolute} and \ref{eq:relative}. Genome-wide micro-averages are $F_q^{\text{OVR}}=\sum_\ell \Delta_q(\ell)/\sum_\ell S_q^{\text{total}}(\ell)$. Analyzing $F_q^{\text{OVR}}$ as a function of $q$ gives a sensitivity spectrum: $q\approx 1$ up-weights rarer alleles (drift/founder signals), and $q\gg1$ emphasizes common alleles ($q=2$ recovers heterozygosity-based separation).

\subsection{Leave-One-Out influence $\Delta F_q$ (LOO): a Population's Contribution to Regional Structure}
Let $K_\mathcal{R}=|\mathcal{R}|$ be the number of populations in region $\mathcal{R}$. Using equal population weights $w_k=1/K_{\mathcal{R}}$, define the regional statistic:
\begin{equation}
    F_q^{(\mathcal{R})}=\frac{\sum_\ell \left(S_q(\text{Bern}(\bar{p}_\ell)) - \frac{1}{K_\mathcal{R}}\sum_{k\in\mathcal{R}} S_q(\text{Bern}(p_{k\ell})) \right)}{\sum_\ell S_q(\text{Bern}(\bar{p}_\ell))}
\end{equation}
Define the LOO influence of population $c\in\mathcal{R}$ as:
\begin{equation}
    \Delta F_q(c)\coloneqq F_q^{(\mathcal{R})}-F_q^{(\mathcal{R} \setminus \{c\})}
\end{equation}
with interpretation such that $\Delta F_q(c)>0$ means removing $c$ reduces between-population differentiation (so $c$ is a driver of structure), and $\Delta F_q(c) <0$ means $c$ homogenizes the region.

\section{Experiments and Discussion}

\subsection{Entropy Analysis of Haplotypes across the Pacific}
We analyze Oceanian and Southeast Asian groups partitioned into Polynesia, Micronesia, Melanesia, and Southeast Asia (Figure \ref{fig:pacific_groups}.C), using 865 samples from the Ocenian dataset \cite{quinto2024genomic}, each with 1,823,000 biallelic SNPs. We split diploid genomes into a resulting dataset of 1,730 haplotypes. As sample sizes vary by population, we report equal-weight estimates by bootstrapping haplotypes within each population (100 resamples, with a per-population cap of 40) and aggregating across loci. This avoids confounding $F_q^{\text{OVR}}$ and $\Delta F_q^{\text{LOO}}$ signals with sample size.

\begin{figure}[ht!]
    \centering
    \includegraphics[width=1.0\linewidth]{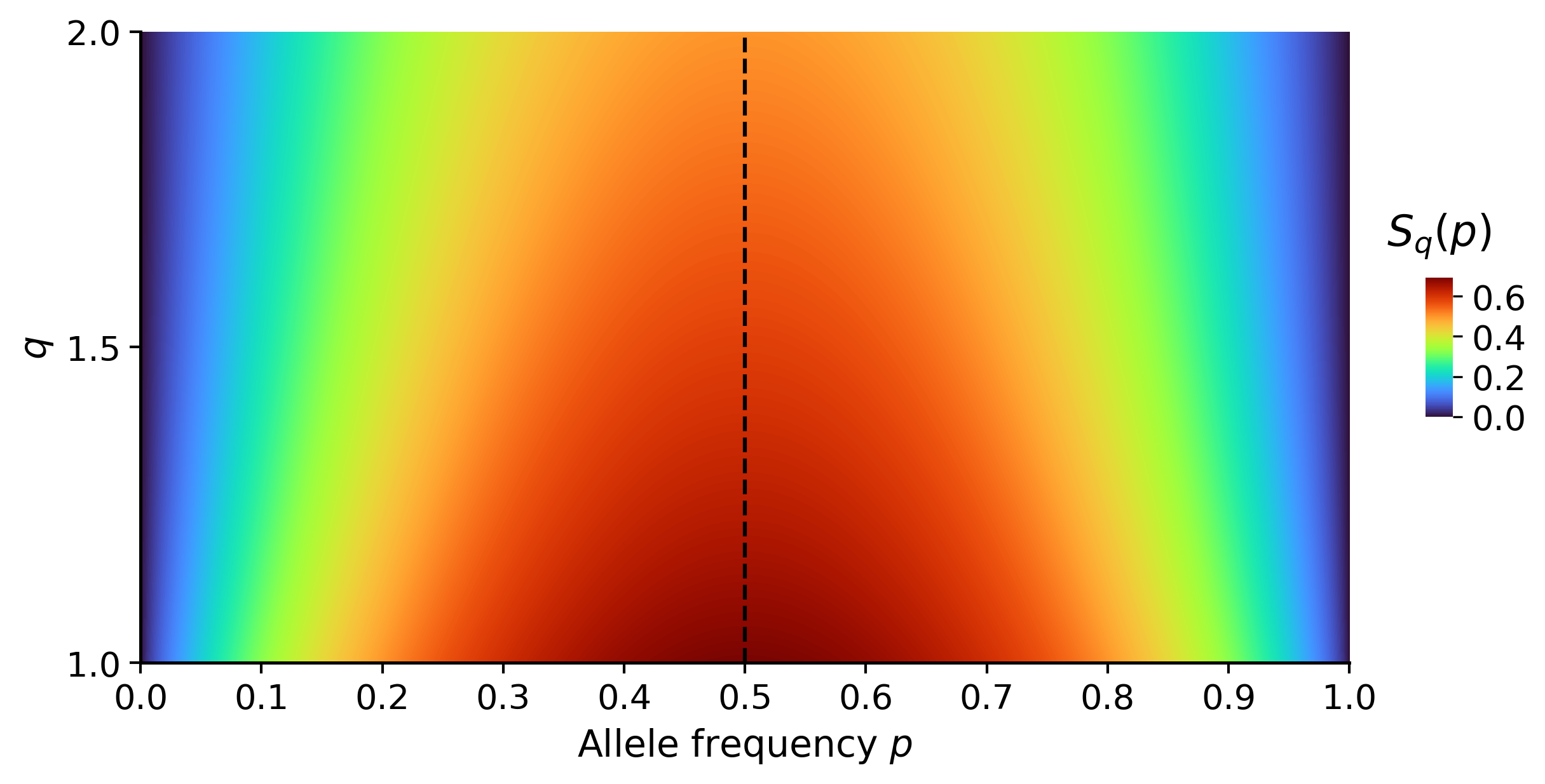}
    \caption{\textbf{Tsallis entropy across allele frequencies and values of $\bm{q}$}. Heatmap of $S_q(p)$ for a biallelic locus, with allele frequency $p\in[0,1]$ and $q\in[1,2]$.}
    \label{fig:entropy_plot}
\end{figure}

Note that in the range $q {\in}[1,2]$, higher $q$ up-weights common variants, while lower $q$ up-weights rarer variants (Figure \ref{fig:entropy_plot}). Thus, it is interesting to observe the slope of $F_q^{\text{OVR}}$ as it decreases (as $q$ increases): a steeper drop from $q=1$ to $q=2$ signals recent drift or serial founder effects, whereas a flatter $F_q^{\text{OVR}}$  profile points to older structure dominated by common variants. This is exactly the tension we want to visualize across islands with different peopling histories.

\subsubsection{Polynesia}
\begin{figure}[ht!]
    \centering
    \includegraphics[width=1.0\linewidth]{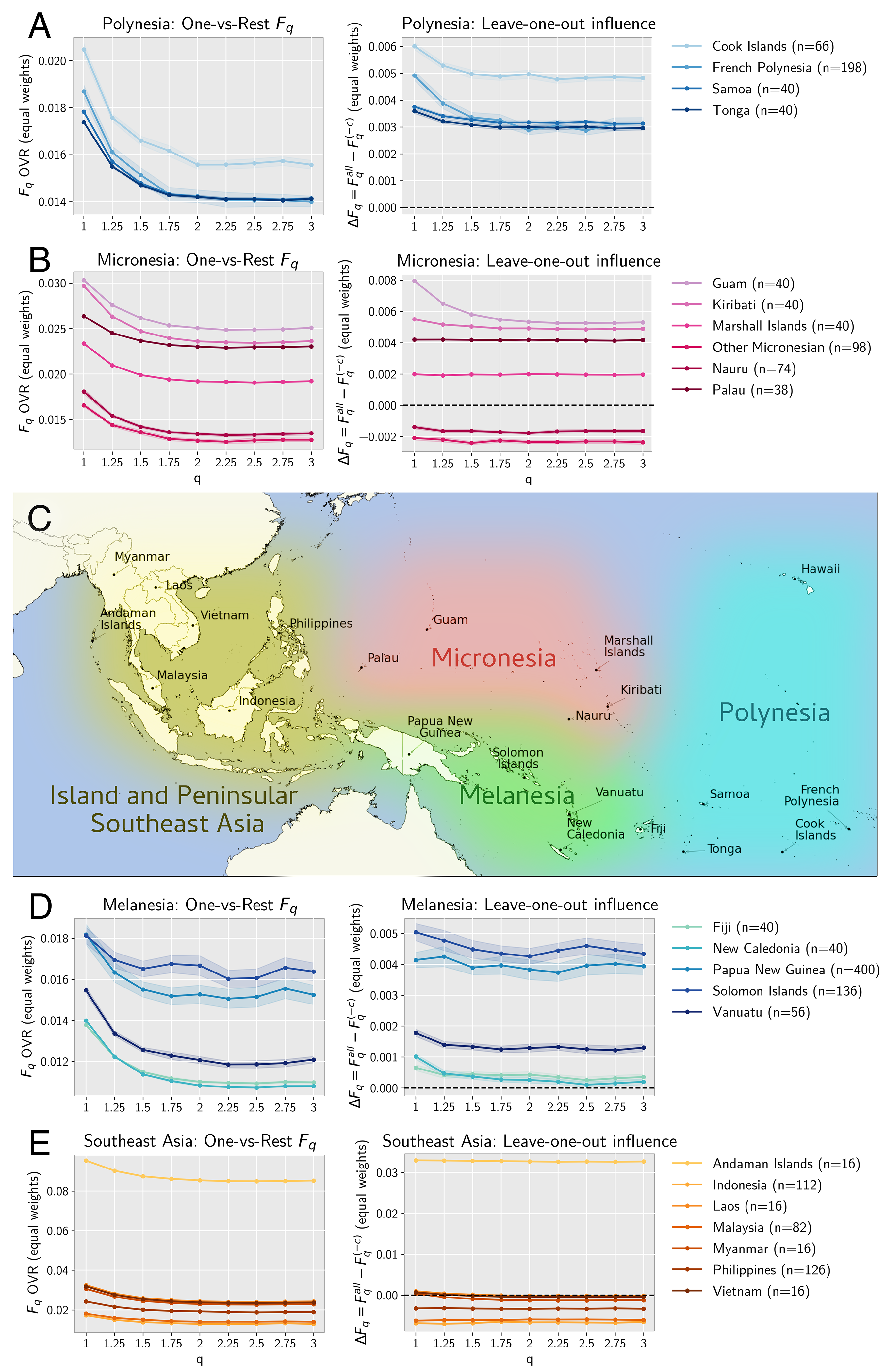}
    \caption{\textbf{Regional differentiation profiles $\bm{F_q}$ (OVR) and $\bm{\Delta F_q}$ (LOO) across the Pacific and Southeast Asia.}
\textbf{(A, B, D, E):} One–vs–Rest $F_q$ (left of each pair) and leave–one–out influence $\Delta F_q^{\text{LOO}}$ (right of each pair) for \emph{Polynesia}, \emph{Micronesia}, \emph{Melanesia}, and \emph{Southeast Asia}. The counts correspond to haplotype counts, twice the number of individuals). Lines show Tsallis $q$-entropy colored by population; shaded ribbons are bootstrap 95\% CIs from resampling. Equal–country weighting is used within each macroregion to reduce sample–size imbalance. \textbf{(C):} locator map with sampling sites (black points) and region polygons used for grouping.}

    \label{fig:pacific_groups}
\end{figure}

The consensus peopling model for East Polynesia (Cook Islands, French Polynesia) features rapid, late settlement following a serial founder expansion from West Polynesia (Samoa, Tonga) \cite{ioannidis2021paths, wilmshurst2011high}. %Our results track that narrative: 
Cook Islands show the largest positive $\Delta F_q^{\text{LOO}}$ across $q$ (Figure \ref{fig:pacific_groups}.A), marking them as especially differentiating within the Polynesian set (i.e., genetically distinct) likely due to continental admixture. %Samoa and Tonga have smaller, positive $\Delta F_q^{\text{LOO}}$, consistent with their more central position within Polynesian variation along well-documented west-to-east dispersal corridors \cite{ioannidis2021paths}. 
French Polynesia follows a similar pattern at high $q$ as Samoa and Tonga, they are differentiated at rare marker variance ($q=1$); this may be one of the effects of French Polynesia having experienced more founder effects \cite{ioannidis2021paths}.

\subsubsection{Micronesia}
Micronesia's history is complex and heterogeneous, with five migratory streams of ancestry and variable Near-Oceania (Papuan-related) contributions \cite{liu2022ancient}, including Melanesian and Polynesian. 
In our panel, Guam, Kiribati, and Palau rank consistently high in OVR and show clearly positive LOO influence, whereas Nauru has the lowest, with a small negative $\Delta F_q^{\text{LOO}}$ (Figure \ref{fig:pacific_groups}.B), acting as a homogenizer in the regional mixture. This reinforces \cite{koji2000mtdna} which found Nauru to be genetically clustered together with other Micronesian populations, while Guam and Palau lie further since they have been found to have more East-Asian like ancestry \cite{liu2022ancient}.

\subsubsection{Melanesia}
Within Melanesia we observe (Figure \ref{fig:pacific_groups}.D) elevated $F_q^{\text{OVR}}$ for Near-Oceanian islands (Papua New Guinea and the Solomon Islands), while Remote Oceania islands (Fiji and New Caledonia) show to have the lowest OVR. This gradient mirrors the findings by \cite{choin2021genomic, skoglund2016genomic} that found Vanuatu and other Remote Oceanian populations have an admixed background from gene flow coming from admixed Near Oceanian (Papua New Guinea and Solomon Islands) individuals. 
% this is consistent with previous finding with later introgression of Papuan ancestry into Remote Oceania \cite{skoglund2016genomic}. 

\subsubsection{Southeast Asian Islanders and Neighboring Coastal Populations}
Finally, in the Southeast Asia region, Andaman Islands appear high on both $q{=}1$ and $q{=}2$ relative to the mainland Southeast Asian populations (Figure \ref{fig:pacific_groups}.E), in line with well-documented isolation and strong drift in these small island populations \cite{sitalaximi2023genetic, thangaraj2006unique}. In contrast, mainland groups such as Myanmar, Laos, and Vietnam behave as regional intermediates, and contribute less to between-population variance with lower LOO influence in our set.

\subsection{Tsallis-Entropy Analysis of Simulated Haplotypes Across Generations}
We evaluate $F_q$ on controlled simulations seeded with real African whole genomes. Founders were drawn from high-coverage HGDP \cite{cann2002human} and 1000 Genomes \cite{the_1000_genomes_project_consortium_global_2015} datasets; variants were selected across autosomes and filtered to 322,216 biallelic SNPs. To avoid cryptic relatedness, we remove pairs up to third-degree using KING kinship inference software \cite{manichaikul2010robust}. The final reference panel contains 1,432 unrelated individuals. Sex-specific recombination is modeled with refined male and female genetic maps \cite{bherer2017refined} to better match crossover landscapes observed in human meiosis. References are clustered into three broad African demes by genetic and ethnolinguistic affinity, namely: \emph{West Africa} (\textsf{WA}), \emph{East Africa} (\textsf{EA}), and \emph{Central, Southern, and Northern Africa} (\textsf{CSN}). 

\subsubsection{Simulation Design} We simulate 17 generations with monogamous pairings. Candidate mates are pruned by sex and kinship rules: cousins closer than second degree are disallowed (including \emph{removed} variants, e.g., 1C1R), and among first cousins only cross-cousin unions are permitted \cite{kirby2016d}. Offspring follow $\text{Poisson}(\lambda=3)$. Between-deme mating is governed by a \emph{panmixia} parameter $\rho\in[0,1]$: $\rho=0$ corresponds to strict endogamy and $\rho=1$ to strictly exogamous random pairing across demes. 

\subsubsection{Baseline Drift (Piecewise Panmixia)} 
\begin{figure}[ht!]
    \centering
    \includegraphics[width=1\linewidth]{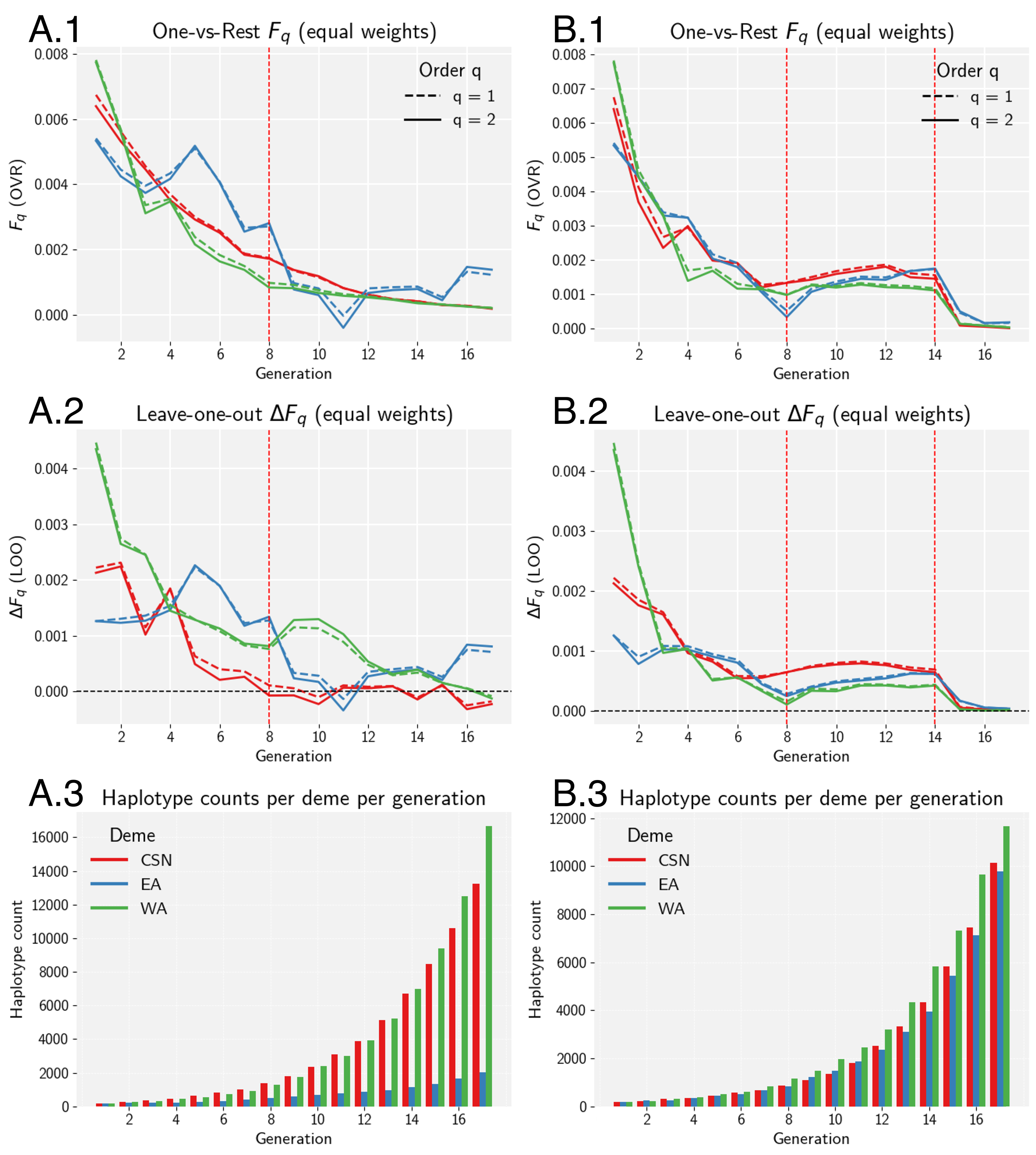}
    \caption{\textbf{Time–series behavior of $\bm{F_q}$ (OVR) and $\bm{\Delta F_q}$ (LOO) under controlled mating policies.}
\textbf{(A.1, B.1):} One–vs–Rest (OVR) $F_q$ across generations for three demes (\textsf{WA}, \textsf{EA}, \textsf{CSN}); solid lines: $q{=}2$ (heterozygosity/second–order), dashed: $q{=}1$ (Shannon/first–order). \textbf{(A.2, B.2):} Leave–one–out (LOO) influence $\Delta F_q^{\text{LOO}}$, measuring each deme's contribution to between–deme differentiation. \textbf{(A.3, B.3):} Haplotype sample counts per deme generated at each generation. \textbf{(A.3):} \emph{Baseline drift} with deme–specific random–mating probabilities $\rho$ changed at generation 8 (red dashed line) from $(\rho_{\textsf{WA}},\rho_{\textsf{EA}},\rho_{\textsf{CSN}})=(0.3,0.5,0.1)$ to $(0.1,0.6,0.3)$. \textbf{(B.3):} \emph{Isolation–reconnection} with $\rho$ set to $(0.5,0.5,0.5)$ initially, then near–isolation $(0.05,0.05,0.05)$ at generation 8 and strong exogamy $(0.9,0.9,0.9)$ at generation 14 (red dashed lines). Curves are genome–wide micro–averages with equal weights across demes.}
\label{fig:simulations}
\end{figure}

We initialize our panmixia policy as $(\rho_{\textsf{WA}},\rho_{\textsf{EA}},\rho_{\textsf{CSN}})=(0.3,0.5,0.1)$. At generation 8, we isolate \textsf{WA}, increase \textsf{EA} mixing, and relax \textsf{CSN}, yielding  $(0.1,0.6,0.3)$. Observe in Figure \ref{fig:simulations} (Column A) that $F_q^{\text{OVR}}$ decays fast early as unrelated founders begin interbreeding. $\Delta F_q^{\text{LOO}}$  
pinpoints which deme carries the structure at each time: prior to gen-8, \textsf{EA} shows elevated $\Delta F_q^{\text{LOO}}$ despite being the most panmictic: \textsf{EA}'s higher $\rho$ turns it into a source of migrants, so removing \textsf{EA} lowers between-deme differentiation. After gen-8, \textsf{WA}'s  $\Delta F_q$ rises while \textsf{EA}'s flattens---consistent with \textsf{WA} becoming more endogamous ($\rho=0.1$) and thus more distinctive, whereas \textsf{EA}'s extra mixing ($\rho=0.6$) dilutes its contrast, but is restored at gen-12 due to the source-sink dynamics. % i.e., their population keep a small growth  and stands out because WA and CSN mix within each other with EA's genetic footprint. 
Overall, under gradual drift with changing mating regimes, OVR tracks the level of \emph{differentiation}, while LOO attributes \emph{responsibility} for structure.

\subsubsection{Isolation-Reconnection Pulse} We define a policy where we start from moderate flow for all demes, $(\rho_{\textsf{WA}},\rho_{\textsf{EA}},\rho_{\textsf{CSN}})=(0.5,0.5,0.5)$. At generation 8, we enforce near-isolation with panmixia set to $(0.05, 0.05, 0.05)$. At generation 14, we flip to strong exogamy $(0.9, 0.9, 0.9)$. Observe in Figure 2 (Column B) that after the beginning of isolation at gen-8, both $F_q^{\text{OVR}}$ and $\Delta F_q^{\text{LOO}}$  rise together in each deme, reflecting classic drift under isolation. Immediately after gen-14, once $\rho{=}0.9$ reconnects demes with intensive migration, both curves drop sharply across demes as the pool re-homogenizes. As a practical implication, for simulator audits or real time-series, the combination of OVR (level) and LOO (attribution) distinguishes which population is structurally distinct, when isolation begins, and when contact resumes.

\begin{comment}
\begin{table}[h]
\caption{\textbf{Demes.}}
\label{tab:demes}
\centering
\begin{tabularx}{\linewidth}{ccc}
\toprule
\textbf{Deme} & \textbf{Number of HGDP samples} & \textbf{Populations included} \\
\midrule
West Africa (WA) & 449 & Yoruba, Esan, Mende, Mandinka \\
East/Horn and Great Lakes (EA) & 108 & Luhya, Luo, Maasai, Dinka, Somali, Bantu Kenya \\
Central/Southern and North Africa (CSN) & 91 & Biaka, Mbuti, San, Khomani San, Bantu South Africa, Bantu Tswana, Herero, Bantu Herero, Tswana, Mozabite, Saharawi
\\
\bottomrule
\end{tabularx}
\end{table}
\end{comment}

\section{Conclusion}
Variance-based fixation metrics---epitomized by $F_{\text{ST}}$, remain foundational but are most sensitive to common variants and can underweight drift signals when the spectrum is skewed. For simulation realism, this implies that simulators must reproduce not just variance-based divergence, but also rare-variant behavior. $F_q$ reframes fixation through entropy: $q{=}1$ (Shannon entropy) emphasizes low-frequency variants, while $q{=}2$ matches standard heterozygosity $F_{\text{ST}}$. For practice, we propose reporting OVR and LOO: viewing OVR curves accross $q$ yields a slope diagnostic, where a steep drop from $q=1$ to $q=2$ flags recent drift or serial founder effects, and a flat profile implies older structure supported by common variants or ongoing gene flow. This complements, rather than replaces, $F_{\text{ST}}$ and related measures. 

\subsection*{Funding Acknowledgments}
No funding was received for this study. The authors have no relevant financial or nonfinancial interests to disclose.

\subsection*{Compliance with Ethical Standards statement} 
We used published data from \cite{the_1000_genomes_project_consortium_global_2015, cann2002human, quinto2024genomic}, each of which had ethical approval through their respective IRBs documented in each reference. No additional ethical approval was required.

\bibliographystyle{IEEEtran}
\bibliography{bibiliography}

\end{document}